\pdfoutput=1
\documentclass[prb,aps,showpacs,twocolumn,preprintnumbers,
amsmath,amssymb,superscriptaddress,longbibliography]{revtex4-2}
\usepackage[english]{babel}
\usepackage{amsmath,amssymb,amsfonts}
\usepackage{graphicx}
\usepackage[colorlinks=true,linkcolor=red,citecolor=blue,urlcolor=blue]{hyperref}

\usepackage{booktabs}
\usepackage[dvipsnames]{xcolor}
\usepackage{braket}
\usepackage{bm}
\usepackage{bbm}
\usepackage{enumitem}
\usepackage{float}
\usepackage{multirow}
\usepackage{longtable}
\usepackage[normalem]{ulem}
\usepackage{array}
\usepackage{makecell}
\usepackage{subfigure}
\graphicspath{{NewFigures/}}
\usepackage{dcolumn}
\newcolumntype{C}{>{$}c<{$}}

\usepackage{listings}

\definecolor{dkgreen}{rgb}{0,0.6,0}
\definecolor{gray}{rgb}{0.5,0.5,0.5}
\definecolor{mauve}{rgb}{0.58,0,0.82}

\allowdisplaybreaks
\DeclareMathAlphabet{\zc}{OT1}{pzc}{m}{it}

\begin{document}

\title{Layer-Resolved Topological Metals in the Bilayer Lieb Lattice}

\author{Mengjie Yang}\email{mengjie.yang@u.nus.edu}
\affiliation{Department of Physics, National University of Singapore, Singapore 117551, Singapore}

\author{S Rahul}
\affiliation{Department of Physics, Manipal Institute of Technology Bengaluru, Manipal Academy of Higher Education, Manipal-576104, India}

\author{Giandomenico Palumbo}
\affiliation{CFisUC, Department of Physics, University of Coimbra, Rua Larga, 3004-516 Coimbra, Portugal}

\date{\today}

\begin{abstract}
We identify a two-dimensional time-reversal-invariant topological metallic phase on a bilayer Lieb lattice, characterized by a quantized layer--resolved pseudo-spin Chern number. Without the orbital-angular-momentum-dependent (OAM-dependent) coupling, the system gives rise to a time-reversal-invariant topological semimetal with a zero indirect gap and quantized pseudo-spin Chern number. Opposite-sign intralayer OAM-dependent coupling immediately converts the zero-indirect-gap semimetal into a metal, in which the global spectrum is metallic while the layer--resolved pseudo-spin Chern number remains well defined as long as the direct gap at each crystal momentum and the pseudo-spin gap remain open. The model also exhibits asymmetric boundary states: in the semimetallic regime, one edge hosts perfectly flat bands, whereas the opposite edge supports gapless counter-propagating modes forming a one-dimensional Dirac cone. An edge-localized interlayer coupling gaps only the counter-propagating edge states, leaving the flat-band edge essentially intact, while intralayer OAM-dependent coupling bends the exact flat band into a dispersive boundary mode without affecting the gapped Dirac edge. These results open a route toward the controlled engineering of layer--resolved topological gapless phases in synthetic and quantum materials.
\end{abstract}

\maketitle

\noindent\textit{\bf{Introduction.---}} In recent years, broad classes of topological semimetals \cite{Armitage2018,Gao2019,Lv2021} have been theoretically proposed and experimentally identified in synthetic and solid-state materials. Semimetals are characterized by the absence of a global gap, although they can in general be divided into two main classes: those with nodal structures arising from band-touching points between different bands, such as Weyl \cite{Wan2011,Xu2011,Burkov2011,Bouhon2020} and Dirac semimetals \cite{Young2012,Wieder2016,Palumbo2021} as well as nodal-line \cite{Fang2016,Wieder2019} and nodal-surface semimetals \cite{Wu2018,Salerno2020}, and those with a robust zero indirect gap, for which the bands do not touch yet the maximum of the lower band coincides with the minimum of the upper band. The latter with broken time-reversal symmetry were theoretically predicted in Refs. \cite{Palumbo2015,Juzeliunas2015} in two dimensions and in Refs. \cite{Palumbo2024,Rahul2026} in one dimension. Importantly, some of these phases have recently been supported by experimental evidence \cite{Chen2025,Beitlberger2026}.

Besides semimetallic phases, considerable effort has been devoted to the understanding and characterization of topological metals \cite{Hu2015,Kamenev2018,Bahari2019,Pyrialakos2023,Dong2023,Zhou2024,Zhao2025} and the relative topological classification and bulk-edge correspondence \cite{Cerjan2022,Cheng2023}. Although several two-dimensional time-reversal-invariant topological metals have recently been discussed in the literature \cite{Pan2014,Teppe2016,Xie2021,Xie2023,Cui2025,Liu2026}, here we emphasize the main novelties of our work.

In this paper, we focus on time-reversal-invariant topological semimetallic and metallic phases supporting asymmetric boundary states.
Firstly, we identify a two-dimensional time-reversal-invariant metallic phase on a bilayer Lieb lattice, characterized by a layer-resolved pseudo-spin Chern number, which we refer to as a ``pseudo-spin Chern metal.''
Unlike the gapped spin-Chern insulator previously realized on a bilayer Lieb lattice~\cite{Deng2020AcousticSpinChern} and the direct-gap--protected Chern metal of Ref.~\cite{Pan2014}, our phase originates from a zero-indirect-gap Chern semimetal and retains a quantized layer-resolved invariant across the global gap closing.
In the absence of OAM-dependent coupling, the system realizes a pseudo-spin Chern semimetal: although the spectrum has a zero indirect gap, the relevant occupied subspace remains locally separated and carries a quantized pseudo-spin Chern number.
Secondly, upon turning on opposite-sign intralayer OAM-dependent coupling, we observe a continuous semimetal-to-metal evolution in which the global spectrum becomes metallic through an indirect band overlap. After introducing a projected pseudo-spin operator and corresponding pseudo-spin spectrum in analogy with the spinful systems \cite{sheng2006quantum,prodan2009robustness,Lin2024}, we show that the layer-resolved pseudo-spin Chern invariant remains well defined and quantized as long as the direct gap at each momentum $k$ and the pseudo-spin gap remain open.
Thirdly, we investigate the peculiar boundary states of our model. In the semimetallic regime, under open boundary conditions along the $y$-direction, we show that one edge of the cylinder hosts perfectly flat edge states, whereas the other hosts gapless counter-propagating modes (one-dimensional Dirac cone). By introducing an edge-localized inter-layer coupling, we generate a boundary Dirac mass for these boundary modes. In contrast, the flat edge band at the opposite boundary remains essentially intact.
Notice that although our inter-layer coupling preserves both time-reversal and chiral symmetry and the single gapped edge resembles the boundary physics of a higher-order topological metal \cite{Benalcazar2017,Schindler2018HOTI,Kunst2018,Cerjan2022,Liu2024HigherOrderMetal}, the presence of the edge flat band further emphasizes the asymmetric nature of our boundary states.
Moreover, intralayer OAM-dependent coupling drives an edge evolution from flat to dispersive, bending the initially exact flat band into a dispersive boundary mode, while the boundary gapped Dirac cone remains unchanged.
Our work opens a route toward the controlled engineering of layer-resolved topological gapless phases with asymmetric boundary states in synthetic and solid-state materials.

\begin{figure}[!htp]
  \centering
  \includegraphics[width=0.95\linewidth]{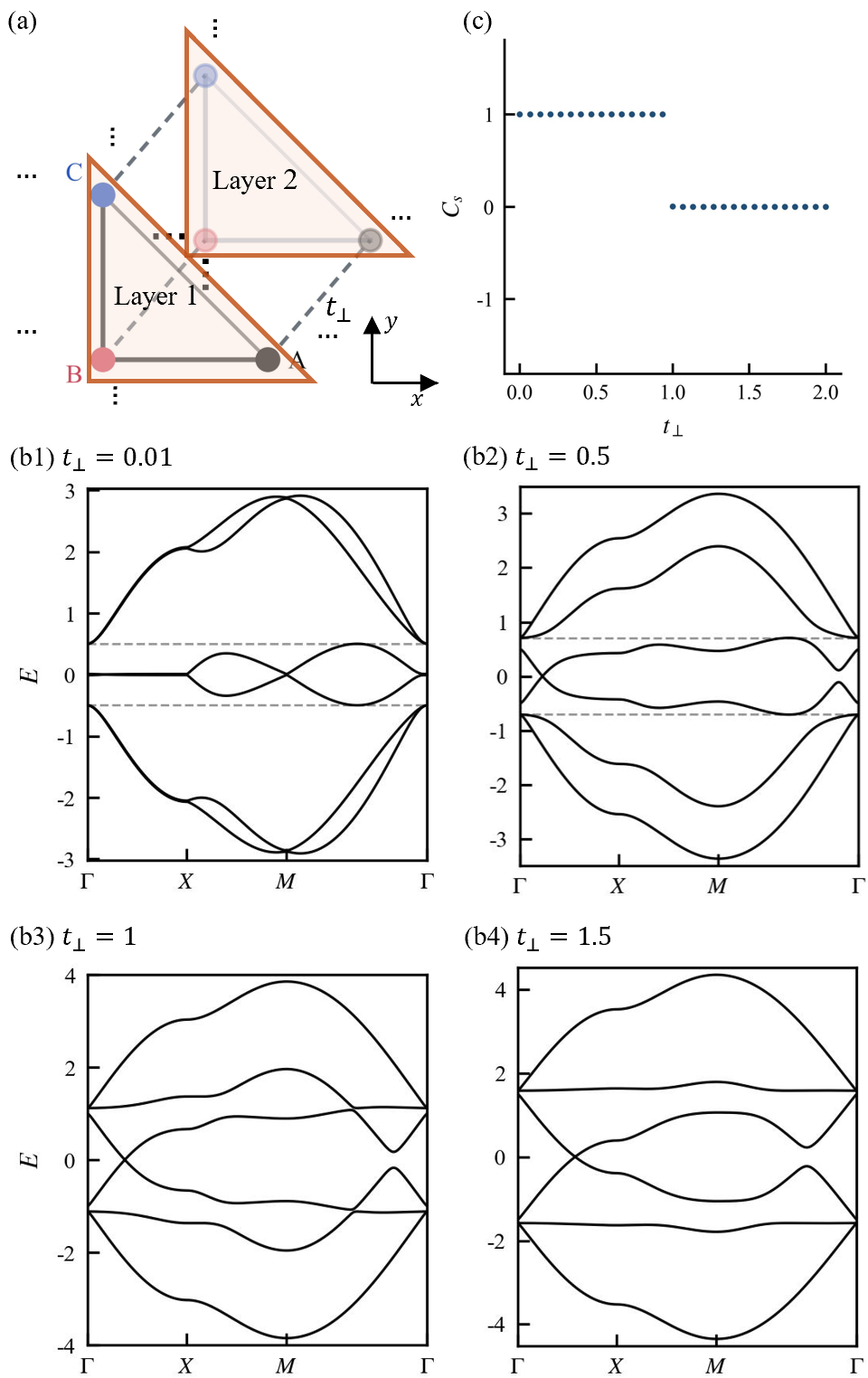}
  \caption{Bulk OAM-coupling-free parent spin-Chern semimetal. (a) Bilayer Lieb lattice for the parent Hamiltonian $H_0(\mathbf{k})$ in Eq.~\eqref{eq:bilayer_parent}, built from the Chern-semimetal block in Eq.~\eqref{eq:single_layer}. The interlayer coupling is orbital diagonal, $V_{\mathrm{inter}}=t_\perp I_3$. (b1)--(b4) Bulk spectra under periodic boundary conditions, plotted along $\Gamma-X-M-\Gamma$ for representative interlayer couplings $t_\perp=0.01,0.5,1,1.5$. The spectra are plotted in black. In panels (b1) and (b2), the gray dashed horizontal lines mark the energy reference lines used to indicate the zero-indirect gap, which is preserved at small $t_\perp$. (c) Pseudo-spin Chern number $C_s$, defined in Eq.~\eqref{eq:spin_chern}, as a function of $t_\perp$. The nonzero plateau $C_s=1$ identifies the time-reversal-invariant pseudo-spin Chern semimetallic phase, while the change to $C_s=0$ at larger $t_\perp$ indicates the topologically-trivial phase.}
  \label{fig:fig1}
\end{figure}

\noindent\textit{\bf{Pseudo-spin Chern semimetal.---}}
We start considering a tight-binding model of spinless fermions without OAM-dependent coupling on the bilayer Lieb lattice with sublattices $(A,B,C)$ as shown in Fig.~\ref{fig:fig1}(a). On each layer, we consider a Chern semimetal \cite{Palumbo2015} such that its corresponding Bloch Hamiltonian is given by
\begin{equation}
h(\mathbf{k};M)=
\begin{pmatrix}
0 & J+Ke^{ik_x} & M^{*}\\
J+Ke^{-ik_x} & 0 & J+Ke^{ik_y}\\
M & J+Ke^{-ik_y} & 0
\end{pmatrix},
\label{eq:single_layer}
\end{equation}
where $J$ and $K$ are the nearest-neighbor intra- and inter-cell hoppings, respectively. For a purely imaginary $A$-$C$ hopping coupling $M$, the system is characterized by a robust zero-indirect gap and a quantized Chern number that depends on the sign of $M$. We then construct a two-layer parent Hamiltonian by pairing two blocks with opposite Chern numbers
\begin{equation}
H_0(\mathbf{k})=
\begin{pmatrix}
h(\mathbf{k};M) & t_\perp I_3\\
t_\perp I_3 & h(\mathbf{k};-M)
\end{pmatrix},
\label{eq:bilayer_parent}
\end{equation}
where $I_3$ acts in the Lieb-sublattice space and $t_\perp$ is an orbital-diagonal interlayer coupling. Since complex conjugation maps the imaginary hopping coupling as $M\mapsto -M$, the two diagonal blocks are exchanged under a spinless time-reversal operation, namely the parent Hamiltonian satisfies
\begin{equation}
\mathcal{T}H_0(\mathbf{k})\mathcal{T}^{-1}=H_0(-\mathbf{k}),\qquad \mathcal{T}=\tau_x K,
\label{eq:time_reversal}
\end{equation}
with $\mathcal{T}^2=+1$.
Moreover, the parent Hamiltonian is invariant under a chiral symmetry $\mathcal{S}H_0\mathcal{S}^{-1}=-H_0$ with $\mathcal{S}=\tau_y\otimes\mathrm{diag}(1,-1,1)$ and $\mathcal{S}^2=+1$, which is preserved by both $H_{\rm SOC}$ and the boundary term below. Together with $\mathcal{T}^2=+1$ this places the model in class BDI; since the strong two-dimensional BDI index is trivial, the quantized $C_s$ is protected not by the Altland--Zirnbauer class but by the spectral gap of $P\tau_zP$~\cite{prodan2009robustness}.
Throughout the main text, we use an effective two-dimensional Bloch description in which the $\tau$ degree of freedom is treated as an internal, on-site sector within the unit cell: the sector-mixing term $t_\perp I_3$ in Eq.~\eqref{eq:bilayer_parent} is $\mathbf{k}$-independent, and no relative in-plane Bloch phase is attached to $\tau_x$. This is what licenses treating $\mathcal{T}=\tau_xK$ as a genuine local antiunitary operator for the tenfold-way classification above. A literal spatial bilayer, in which the two flavors are hosted on physically stacked layers, is one possible synthetic implementation of this internal-sector model (see Sec.~S3), but not the only one, as we discuss next.

Indeed, $H_0(\mathbf{k})+H_{\rm SOC}(\mathbf{k})$ admits an equally valid reading as a spinless three-orbital (Lieb) model in which the $\tau=\pm$ sectors correspond to an on-site $p_x\pm ip_y$ orbital doublet ($L_z=\pm1$), with $M\mapsto-M$ under complex conjugation reflecting $L_z\to-L_z$ under time reversal; in this language $H_{\rm SOC}$ is an orbital-angular-momentum-dependent (OAM-dependent) sublattice coupling rather than a spin-dependent one. This construction is closely related to the feature/orbital-resolved topological framework developed in Refs.~\cite{Wang2023FeatureSpectrum,Yao2024FeatureEnergy,Wang2025OrbitalHall}, where a projected internal-degree-of-freedom spectrum ($P\hat O P$, here $\hat O=\tau_z$) is used to define sector-resolved Chern numbers and their associated bulk-boundary correspondence. In particular, the orbital Hall framework of Ref.~\cite{Wang2025OrbitalHall} uses a projected OAM spectrum to define orbital-resolved Chern numbers and the corresponding bulk-boundary correspondence.
The bulk spectrum of the bilayer system with $J=-1$, $K=1$, $M=0.5i$ and different values of $t_\perp$ is shown in Figs.~\ref{fig:fig1}(b1-b4).
For small values of $t_\perp$, the system is still characterized by a zero indirect gap, as in the original Chern semimetal \cite{Palumbo2015}.
Within this regime, we can now define a quantized topological invariant.
Since the two layers carry opposite Chern numbers, the total Chern number vanishes, but the projected layer sectors carry a quantized pseudo-spin Chern number
\begin{equation}
C_s=\frac{1}{2}(C_+-C_-),
\label{eq:spin_chern}
\end{equation}
where $C_\pm$ are obtained from the positive and negative eigenvalue sectors of $P\tau_zP$ using the lattice formula for the Chern number~\cite{FukuiHatsugaiSuzuki2005}. Fig.~\ref{fig:fig1}(c) numerically shows that $C_s=1$ over the parameter regime for which the zero indirect gap is preserved. This construction is structurally identical to the feature-spectrum decomposition used to define orbital- or spin-resolved Chern numbers in Refs.~\cite{Wang2023FeatureSpectrum,Yao2024FeatureEnergy,Wang2025OrbitalHall}.

\begin{figure}[!htp]
  \centering
  \includegraphics[width=0.95\linewidth]{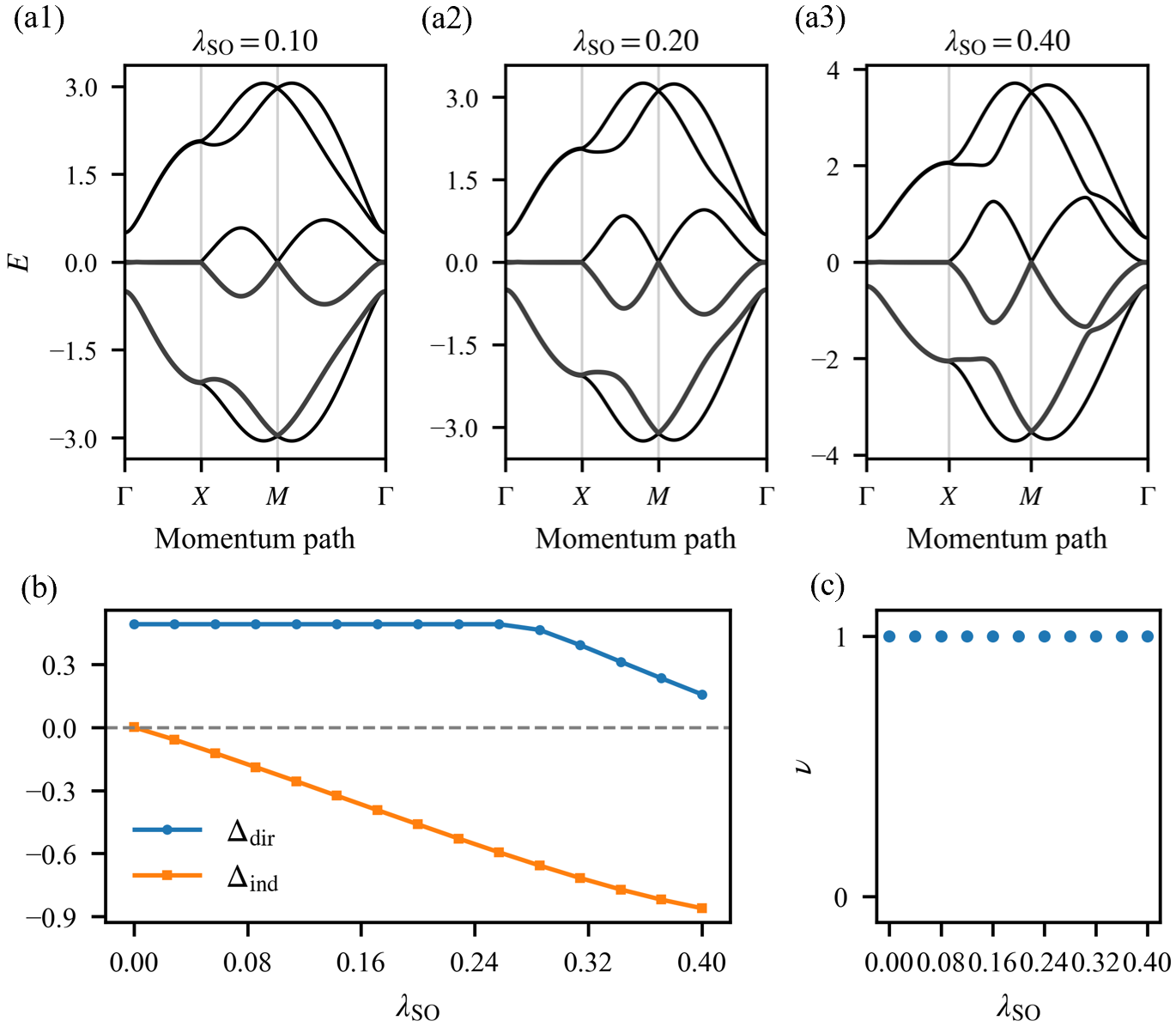}
  \caption{Bulk spectrum and parity diagnostic of the direct-gap-protected pseudo-spin Chern metal with opposite-sign intralayer $\tau_z$ OAM-dependent coupling (labeled $\lambda_{\rm SO}$ for consistency with the axes; see the footnote after Eq.~\eqref{eq:soc}), Eqs.~\eqref{eq:soc} and \eqref{eq:km_block}. (a1)--(a3) Bulk band structures along $\Gamma$--$X$--$M$--$\Gamma$ for $\lambda_{\mathrm{SO}}=0.10,0.20,0.40$. The OAM-dependent coupling term deforms the central bands and drives an indirect band overlap, while the lower two-band subspace remains separated from the remaining bands by a finite direct gap in the plotted range. (b) Direct and indirect gaps, defined in Eq.~\eqref{eq:direct_indirect_gap}, as functions of $\lambda_{\mathrm{SO}}$, evaluated over the full Brillouin zone. The regime with $\Delta_{\mathrm{dir}}>0$ and $\Delta_{\mathrm{ind}}<0$ corresponds to a direct-gap-protected metallic regime. (c) Wilson-loop parity diagnostic $\nu=C_s \bmod 2$ for the lower two-band projector. This parity should be understood as a diagnostic of the projected pseudo-spin Chern structure, not as the Fu--Kane invariant of a $T^2=-1$ quantum spin Hall system.}
  \label{fig:fig2}
\end{figure}

\noindent\textit{\bf{Pseudo-spin Chern metal.---}}
The model in the previous section is a time-reversal-invariant topological semimetal. We now ask whether its topology survives once it is converted into a metal. To this end, we switch on the opposite-sign intralayer OAM-dependent coupling and track its bulk evolution in Fig.~\ref{fig:fig2}.
In the bilayer basis, our OAM-dependent coupling term is
\begin{equation}
H_{\rm SOC}(\mathbf{k})=\lambda_{\rm SO}\tau_z\otimes h_{\rm AC}(\mathbf{k}),
\label{eq:soc}
\end{equation}
\footnote{We retain the subscript ``SO'' in $H_{\rm SOC}$ and $\lambda_{\rm SO}$ as a label for consistency with all figures. Physically, this term is an orbital-angular-momentum-dependent (OAM-dependent) sublattice coupling rather than genuine spin--orbit coupling, since $\tau_z$ labels a layer/orbital pseudospin, not real electron spin (see the remark below Eq.~\eqref{eq:time_reversal}).}
where $h_{\rm AC}$ only couples the $A$ and $C$ sublattices,
\begin{equation}
\begin{aligned}
[h_{\rm AC}(\mathbf{k})]_{AC}&=g(\mathbf{k}),\\
[h_{\rm AC}(\mathbf{k})]_{CA}&=g^{*}(\mathbf{k}),
\end{aligned} \label{eq:km_block}
\end{equation}
$g(\mathbf{k})=-i(1-e^{ik_x})(1-e^{-ik_y}). $
The form factor obeys $g^{*}(\mathbf{k})=-g(-\mathbf{k})$, and hence the full sublattice matrix satisfies
$
h_{\rm AC}^{*}(\mathbf{k})=-h_{\rm AC}(-\mathbf{k}).
$
Using this relation together with $\tau_x\tau_z\tau_x=-\tau_z$, one obtains
$
\mathcal{T}H_{\rm SOC}(\mathbf{k})\mathcal{T}^{-1} =\lambda_{\rm SO}(\tau_x\tau_z\tau_x)\otimes h_{\rm AC}^{*}(\mathbf{k}) =\lambda_{\rm SO}\tau_z\otimes h_{\rm AC}(-\mathbf{k}) =H_{\rm SOC}(-\mathbf{k}).$
Thus, the full Hamiltonian
\begin{equation}
H(\mathbf{k})=H_0(\mathbf{k})+H_{\rm SOC}(\mathbf{k})
\end{equation} preserves the time-reversal symmetry in Eq.~\eqref{eq:time_reversal}. Since $\tau_z$ is diagonal in layer space, the two layers carry opposite-sign intralayer OAM-dependent coupling.

\begin{figure*}
  \centering
  \includegraphics[width=0.95\linewidth]{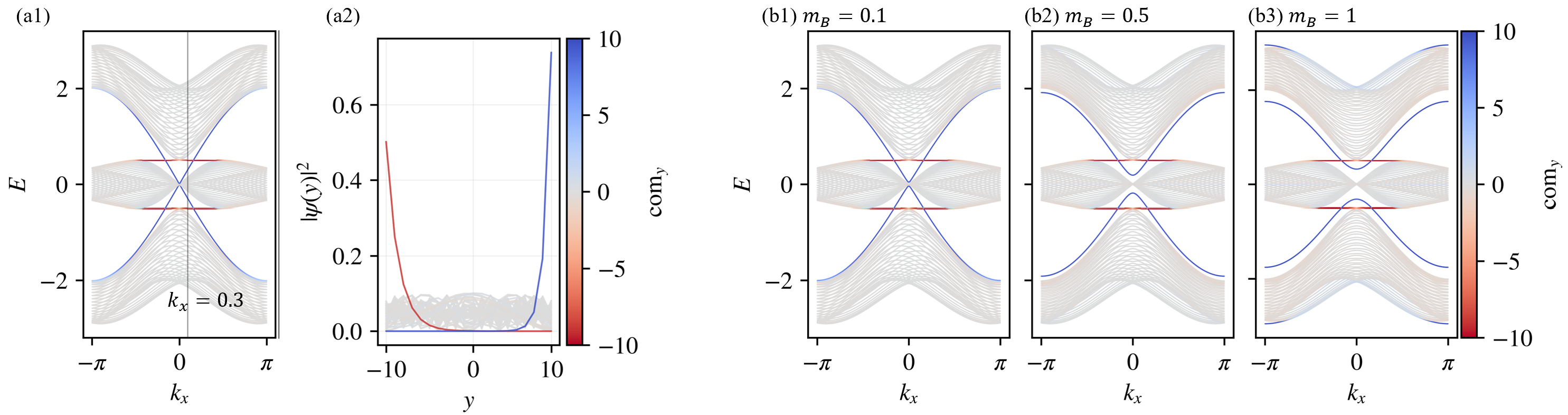}
  \caption{Edge states for the Hamiltonian in Eq.~\eqref{eq:bilayer_parent} at $\lambda_{\mathrm{SO}}=0$. (a1) $m_B=0$. Energy spectrum $E(k_x)$ with periodic boundary conditions along $x$ and open boundary conditions along $y$. The grey curves denote the full finite-strip spectrum, while the highlighted red and blue branches indicate states localized near opposite $y$ boundaries, classified by their center of mass $\mathrm{com}_y$. The vertical line marks the representative momentum $k_x=0.3$ used in (a2). (a2) Real-space probability profiles $|\psi(y)|^2$ of the selected states at $k_x=0.3$. The red and blue profiles show strong localization at opposite boundaries, whereas the grey profiles correspond to extended or weakly localized strip states. (b1)--(b3) Evolution of the boundary Dirac mass in Eq.~\eqref{eq:boundary_mass}, with $m_B=0.1,0.5,1$. Increasing $m_B$ shifts and gaps the counter-propagating edge states on one edge, while the remaining strip and flat-band states are shown in grey. The color scale denotes $\mathrm{com}_y$, with red and blue corresponding to localization near the two opposite $y$ edges.}
  \label{fig:fig3}
\end{figure*}

\begin{figure*}
  \centering
  \includegraphics[width=0.99\textwidth]{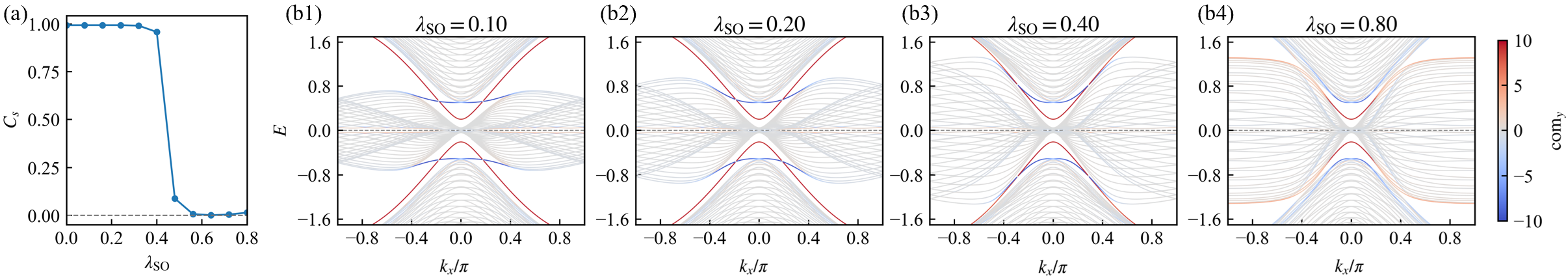}
  \caption{OAM-coupling-driven flat-to-dispersive edge bands and topological phase transition. (a) Cylinder ($x$PBC-$y$OBC) pseudo-spin Chern marker $C_s$ as a function of $\lambda_{\mathrm{SO}}$ with the edge mass $m_B=0.5$ switched on. $C_s$ stays quantized at $1$ throughout the weak-to-intermediate OAM-coupling range, showing that $H_B$ gaps a single edge without altering the bulk topology of the cylinder, and drops sharply once the direct gap closes near $\lambda_{\rm SO}\simeq0.6$. (b1)--(b4) Ribbon spectra for $\lambda_{\mathrm{SO}}=0.10,0.20,0.40,0.80$, with selected states colored by their transverse center of mass $\mathrm{com}_y$. The gray curves show the full ribbon spectrum, while the red and blue branches correspond to states localized near opposite $y$ boundaries. Around $\lambda_{\mathrm{SO}}\simeq0.60$, the direct gap defined in Eq.~\eqref{eq:direct_indirect_gap} closes, consistent with the abrupt drop to zero of $C_s$ in (a).}
  \label{fig:fig4}
\end{figure*}

As shown in Fig.~\ref{fig:fig2}, the OAM-dependent coupling induces a continuous semimetal-to-metal phase transition: the indirect gap becomes negative, while the relative direct gap between the occupied and unoccupied states remains open at each momentum $k$ over the weak-to-intermediate OAM-coupling regime. Figure~\ref{fig:fig2} shows the spectrum of this topological metallic phase; the survival of the bulk topology across the weak-to-intermediate OAM-coupling range is signaled by the odd Wilson-loop parity $\nu=1$ in Fig.~\ref{fig:fig2}(c) together with the fact that the occupied bands remain locally separated. That this quantization persists with the edge mass switched on is shown, in the cylinder geometry, in Fig.~\ref{fig:fig4}(a) (see Supplemental Material~\cite{suppmat}, Sec. 1).
The precise distinction between the semimetallic and metallic phases can be made explicit by considering the direct and indirect gaps, respectively defined as
\begin{equation}
\begin{aligned}
    &\Delta_{\rm dir}=\min_{\mathbf{k}}\left[E_{n_{\rm occ}+1}(\mathbf{k})-E_{n_{\rm occ}}(\mathbf{k})\right],\\
&\Delta_{\rm ind}=\min_{\mathbf{k}}E_{n_{\rm occ}+1}(\mathbf{k})-\max_{\mathbf{k}}E_{n_{\rm occ}}(\mathbf{k}).
\end{aligned}
\label{eq:direct_indirect_gap}
\end{equation}
Here $n_{\rm occ}=2$ denotes the two lowest bands of the six-band bilayer, the manifold separated from the rest by the direct gap and used to build the projector $P$; the corresponding mixed real-space/momentum-space cylinder marker is defined in Supplemental Material~\cite{suppmat}, Sec. 1.
The regime $\Delta_{\rm dir}>0$ and $\Delta_{\rm ind}<0$ is metallic at the level of the global spectrum, yet the occupied subspace is still separated locally in momentum. Because the lower and middle bands do not touch each other, the values of $C_s$ in the semimetallic phase are preserved in the metallic one. Consistently, the Wilson-loop spectrum in Fig.~\ref{fig:fig2}(c) gives an odd parity diagnostic, $\nu=1$, and the pseudo-spin Chern number remains quantized as long as the direct gap and the projected-pseudo-spin gap related to $P\tau_zP$ do not close \cite{Lin2024}. Moreover, the odd parity diagnostic should be read as $\nu=C_s \bmod 2$, while for strong OAM-dependent coupling, the direct gap closes by destroying the topological phase as shown in Fig.~\ref{fig:fig4}.
We emphasize that, with a negative indirect gap, $C_s$ is a direct-gap--protected band-geometric marker rather than a quantized transport coefficient: the residual Fermi surface precludes a quantized (spin) Hall response~\cite{Beugeling2012}, and the role of $C_s$ is to diagnose the locally separated occupied subspace and the associated in-gap boundary modes, in the spirit of local topological markers for gapless systems~\cite{Cerjan2022,Cheng2023}.
Having established that the bulk topology survives in the presence of OAM-dependent coupling, in the next section, we will focus on the boundary states of our bilayer system.

\noindent\textit{\bf{Boundary states.---}}
To analyze boundary modes, we keep periodic boundary conditions along $x$ and open boundary conditions along $y$, such that the ribbon spectrum directly resolves the two opposite $y$ boundaries (cylinder geometry). As shown in Fig.~\ref{fig:fig3}(a1), in the absence of OAM-dependent coupling, the topological semimetal supports two qualitatively different boundary states: the flat bands localized at one and the dispersive counter-propagating states localized near the opposite edge. Because $\mathcal{T}^2=+1$, this counter-propagating crossing is not Kramers protected and can be gapped by a time-reversal- and chiral-symmetric perturbation, as we show next. We now introduce the following boundary term
\begin{equation}
H_B=m_B\,\tau_x\otimes Q_B\otimes |N_y\rangle\langle N_y|,\qquad
Q_B=\mathrm{diag}(0,1,0),
\label{eq:boundary_mass}
\end{equation}
which preserves both time-reversal and chiral symmetry.
Here, $N_y$ is the rightmost row on $y$, $Q_B$ projects onto the $B$ sublattice, while $|N_y\rangle\langle N_y|$ projects onto the selected open boundary row. Thus, $H_B$ mixes the two layers only on the $B$ sublattice of the right $y$ boundary. Importantly, $H_B$ induces a Dirac-like gap to the dispersive counter-propagating states as shown in Figs.~\ref{fig:fig3}(b1-b3) while preserving the flat bands on the other edge. In this way, we further emphasize the asymmetric nature of the boundary states in our system; a projected massive-Dirac description of the gapped boundary branch is given in Supplemental Material~\cite{suppmat}, Sec. 2.
Notice that this coexistence of gapped modes and flat bands differs from higher- and hybrid-order topological metals or insulators, where gapped first-order boundaries typically coexist with corner or hinge modes instead of flat edge bands~\cite{Benalcazar2017,Schindler2018HOTI,Kunst2018,Kunst2019,Liu2024HigherOrderMetal,Kooi2020HigherOrderWeak,Yang2021HybridOrderPhononic,Zhu2025HybridOrderBoundary}.

As established from the bulk spectra and Wilson-loop parity in the previous section, weak OAM-dependent coupling produces a direct gap giving rise to a metal phase. In the presence of the boundary, the OAM-dependent coupling bends the edge flat bands while remaining continuously connected to the topological ribbon spectrum, as shown in Fig.~\ref{fig:fig4}. Crucially, the cylinder marker $C_s$ in Fig.~\ref{fig:fig4}(a) remains quantized with $m_B=0.5$ on, confirming that the edge-localized mass leaves the bulk region of the cylinder topologically intact; the definition and interpretation of this mixed-geometry projected marker are given in Supplemental Material~\cite{suppmat}, Sec. 1.
Notice that this bending mechanism is similar to what happens to the flat band in the bulk of the Kagome lattice with SOC \cite{Ma2020}.
Moreover, for larger OAM-dependent coupling, around $\lambda_{\rm SO}\simeq0.6$, the ribbon spectrum shows a direct-gap closing between the occupied and unoccupied states, and the pseudo-spin Chern number drops to zero sharply. The loss of quantization is therefore caused by the closing of the direct gap that gives rise to a topological phase transition. We have verified numerically that $\Delta_\tau^{\min}$ (Eq.~\eqref{eq:supp_spin_gap}) remains strictly positive throughout the $C_s=1$ plateau shown in Figs.~\ref{fig:fig1}(c) and \ref{fig:fig4}(a), closing only in the immediate vicinity of the direct-gap closing near $\lambda_{\rm SO}\simeq0.6$ (though we do not show this spectrum separately).

\noindent\textit{\bf{Discussion and outlook.---}}
In this work, we have presented a realization of time-reversal-invariant topological gapless phases characterized by a layer-resolved pseudo-spin-Chern number. The relevant organizing principle is the direct gap: as long as the occupied and unoccupied states remain separated at each momentum, the pseudo-spin Chern diagnostic and its Wilson-loop parity remain meaningful even when the global spectrum is metallic. In this regime, the OAM-dependent coupling converts the topological semimetal into a topological metal and continuously deforms the flat edge bands into dispersive ones.
Moreover, we have shown that it is possible to add a symmetry-preserving boundary term that induces a Dirac gap for the counter-propagating edge modes on one edge while leaving the opposite edge flat bands essentially intact. This produces asymmetric boundary states distinct from the edge modes of conventional or higher-order topological phases~\cite{Liu2024HigherOrderMetal,Kooi2020HigherOrderWeak,Yang2021HybridOrderPhononic,Zhu2025HybridOrderBoundary}.
The strong-OAM-coupling regime sets the limit of this description. Once the direct gap closes due to the OAM-dependent coupling, the occupied states are no longer separated by the unoccupied states giving rise to a topological phase transition.
Possible photonic, topolectrical-circuit, cold-atom, and artificial-electronic implementations, together with the corresponding observables and limitations of the metallic regime, are discussed in Supplemental Material~\cite{suppmat}, Sec. 3.

Several extensions follow from the two features that distinguish this phase: a direct-gap-protected pseudo-spin marker carried by a metallic spectrum, and an asymmetric pair of flat and dispersive boundaries. Non-Hermitian perturbations, such as gain, loss, or nonreciprocal hopping, could test whether the two boundary sectors respond differently to skin accumulation and exceptional-bound-state physics~\cite{YaoWang2018NHChern,LinTaiYang2023TopologicalNHSE,Okuma2023NHReview,YangLee2026EBB,YangLee2026TransverseSwitch}. Disorder and feedback-induced boundary restructuring provide another route to test the robustness of the layer-resolved invariant beyond clean Hermitian ribbons~\cite{YangLee2024Percolation,YangLee2025Feedback}. Interaction effects should also be boundary-selective: the dispersive branch may support counter-propagating Luttinger-liquid physics~\cite{Wu2006HelicalLiquid}, whereas the flat edge band is a natural setting for quantum-geometric and caging-dominated correlated phenomena~\cite{PeottaTorma2015,Torma2022QuantumGeometry,YangYuanLee2025BosonicCaging}. More broadly, the metallic regime invites a sharper comparison between direct-gap-protected pseudo-spin markers and real-space probes of gapless topology~\cite{Cerjan2022,Cheng2023}, and corner-selective perturbations could clarify whether the asymmetric edge structure can coexist with higher-order corner modes~\cite{Liu2024HigherOrderMetal}.

We stress that the boundary states reported in Figs.~\ref{fig:fig3}--\ref{fig:fig4} correspond to one specific choice of edge termination; cutting the ribbon along a different sublattice row, or allowing edge reconstruction, can in general alter the flat-versus-dispersive character of the two boundaries, as is well documented for Lieb-lattice flat bands more generally. Likewise, the quantization of $C_s$ relies on $\Delta_{\rm dir}$ and $\Delta_\tau^{\min}$ remaining open; local disorder that closes either gap, rather than merely broadening the spectrum, would destroy it. In synthetic platforms (photonic lattices, topolectrical circuits) both the termination and the coupling pattern are externally engineered and hence reproducible by construction; in solid-state or cold-atom realizations, boundary-resolved spectroscopy would be needed to confirm which edge structure is realized.

\noindent\textit{\bf{Acknowledgments.---}}
The authors thank Benjamin Wieder for inspiring discussions and careful reading of the manuscript.
G. P. acknowledges support from national funds by FCT - Fundação para a Ciência e Tecnologia, I.P. in the framework of the project UID/04564/2025, with DOI identifier 10.54499/UID/04564/2025.

\clearpage
\newpage
\onecolumngrid

\setcounter{equation}{0}
\setcounter{figure}{0}
\setcounter{table}{0}
\setcounter{section}{0}
\setcounter{page}{1}
\renewcommand{\theequation}{S\arabic{equation}}
\renewcommand{\thefigure}{S\arabic{figure}}
\renewcommand{\thetable}{S\arabic{table}}
\renewcommand{\thesection}{S\arabic{section}}
\renewcommand{\thepage}{S\arabic{page}}

\begin{center}
{\large\bfseries Supplemental Material for}\\[0.4em]
{\large\bfseries ``Layer-Resolved Topological Metals in the Bilayer Lieb Lattice''}\\[1.0em]

{\normalsize Mengjie Yang$^{1}$, S Rahul$^{2}$, and Giandomenico Palumbo$^{3}$}\\[0.6em]

{\small
$^{1}$Department of Physics, National University of Singapore, Singapore 117551, Singapore\\
$^{2}$Department of Science and Humanities, PES University EC Campus, Bangalore 560100, India\\
$^{3}$CFisUC, Department of Physics, University of Coimbra, Rua Larga, 3004-516 Coimbra, Portugal
}
\end{center}

\vspace{1.0em}

\section{Numerical evaluation of the real-space pseudo-spin Chern number}

In this section, we show the topological diagnostic used for the $x$PBC-$y$OBC cylinder geometry. This quantity is not a full two-dimensional open-boundary real-space pseudo-spin Chern number. Instead, it is a projected pseudo-spin-Chern marker: the $x$ direction is treated in momentum space, while the $y$ direction is kept in real space. The construction follows the projected-spin decomposition underlying the spin Chern number~\cite{prodan2009robustness,Lin2024}, and with a local-marker formulation of the pseudo-spin Chern number in a mixed position-momentum representation.\\
For each crystal momentum $k_x$, we diagonalize the ribbon Hamiltonian $H_{\mathrm{cyl}}(k_x)$ with open boundary conditions along $y$. The occupied projector is defined by
\begin{equation}
P(k_x)=\sum_{n=1}^{N_{\mathrm{occ}}}|u_n(k_x)\rangle\langle u_n(k_x)|,
\label{eq:supp_Pkx}
\end{equation}
where $N_{\mathrm{occ}}=n_{\mathrm{occ}}N_y$ for a ribbon of width $N_y$, and $n_{\mathrm{occ}}=2$ in the calculations presented in the main text. We then project the layer-pseudospin operator $\tau_z$ into the occupied subspace,
\begin{equation}
S_P(k_x)=P(k_x)\tau_zP(k_x).
\label{eq:supp_projected_tau}
\end{equation}
When the spectrum of $S_P(k_x)$ is separated into positive and negative sectors for all sampled $k_x$, we define the corresponding spectral projectors
\begin{equation}
P_+(k_x)=\sum_{\xi_a(k_x)>0}|\phi_a(k_x)\rangle\langle\phi_a(k_x)|,\qquad
P_-(k_x)=\sum_{\xi_a(k_x)<0}|\phi_a(k_x)\rangle\langle\phi_a(k_x)|,
\label{eq:supp_sector_projectors}
\end{equation}
where $S_P(k_x)|\phi_a(k_x)\rangle=\xi_a(k_x)|\phi_a(k_x)\rangle$. The minimum projected-pseudo-spin gap is monitored as
\begin{equation}
\Delta_{\tau}^{\mathrm{min}}=\min_{k_x}\min_a|\xi_a(k_x)|.
\label{eq:supp_spin_gap}
\end{equation}
A nonzero $\Delta_{\tau}^{\mathrm{min}}$ is required for the separation into $P_+$ and $P_-$ sectors to be well defined.
This is the quantity underlying the assertion in the main text that $\Delta_\tau^{\min}$ remains open throughout the $C_s=1$ regime.

For each sector $\sigma=\pm$, we evaluate a hybrid Chern marker. In a fully real-space local Chern marker, the Chern number can be written in terms of the projected commutators with the two position operators. In the present cylinder geometry, translational invariance along $x$ allows the replacement
\begin{equation}
[X,P_\sigma]\rightarrow i\partial_{k_x}P_\sigma(k_x),
\label{eq:supp_x_commutator}
\end{equation}
while the $y$ direction remains real-space resolved through
\begin{equation}
[Y,P_\sigma(k_x)]=YP_\sigma(k_x)-P_\sigma(k_x)Y.
\label{eq:supp_y_commutator}
\end{equation}
The sector-resolved hybrid marker is therefore computed as
\begin{equation}
C_\sigma(y)=\frac{2\pi i}{N_k}\sum_{k_x}\mathrm{Tr}_{\alpha}\left\{
P_\sigma(k_x)\left[\partial_{k_x}P_\sigma(k_x),[Y,P_\sigma(k_x)]\right]
\right\}_{y,\alpha;y,\alpha},
\label{eq:supp_hybrid_marker_y}
\end{equation}
where $\mathrm{Tr}_{\alpha}$ denotes the trace over internal orbital and layer degrees of freedom within the unit cell at position $y$. The derivative $\partial_{k_x}P_\sigma(k_x)$ is evaluated numerically using a finite difference on the discretized $k_x$ mesh. The bulk value is obtained by averaging $C_\sigma(y)$ over the central region of the ribbon, excluding a fixed number of sites near the two open boundaries:
\begin{equation}
C_\sigma^{\mathrm{bulk}}=\frac{1}{N_y-2N_{\mathrm{m}}}\sum_{y=N_{\mathrm{m}}+1}^{N_y-N_{\mathrm{m}}}C_\sigma(y),
\label{eq:supp_bulk_average}
\end{equation}
where $N_{\mathrm{m}}$ is the boundary margin used in the numerical calculation. Finally, the hybrid projected spin-Chern marker is defined as
\begin{equation}
C_s^{\mathrm{hyb}}=\frac{C_+^{\mathrm{bulk}}-C_-^{\mathrm{bulk}}}{2}.
\label{eq:supp_hybrid_spin_chern}
\end{equation}
This marker should be interpreted as a cylinder-geometry diagnostic rather than as an independent bulk invariant. It uses the same projected-spin sector decomposition as the bulk spin Chern number, but it is evaluated in the same $x$PBC-$y$OBC geometry as the ribbon spectra. This is useful for checking whether the central region of the finite ribbon retains the layer-resolved pseudo-spin Chern number after the boundary perturbation is introduced. In particular, in Fig.~4 the edge-localized interlayer mass $m_B$ modifies the open-boundary spectrum but does not modify the bulk topology. The role of $C_s^{\mathrm{hyb}}$ is therefore to show that, away from the perturbed edge, the bulk region of the cylinder remains consistent with the quantized projected-spin topology as long as the direct band gap and the projected-pseudo-spin gap do not close.

\section{Effective boundary Dirac theory and anomalous half winding}

In this section we provide a low-energy interpretation of the boundary mass term induced by the edge-localized interlayer coupling used in the main text. The purpose is not to define a new bulk invariant, but to show that the gapped dispersive boundary branch is described after projection by a single massive Dirac edge theory. This continuum edge theory carries the anomalous half-integer winding contribution associated with a single massive Dirac fermion.

We consider the ribbon geometry with periodic boundary conditions along $x$ and open boundary conditions along $y$ in the OAM-coupling-free parent. In the absence of the edge-localized interlayer coupling, the dispersive edge branch near $k_x=0$ is locally described by a massless Dirac Hamiltonian in the two-dimensional subspace spanned by the counter-propagating boundary modes,
\begin{equation}
H_{\mathrm{edge}}^{(0)}(k_x)=v k_x\sigma_x,
\end{equation}
where $\sigma_\mu$ acts in the effective edge-mode subspace and $v$ is the edge velocity. The edge-localized interlayer perturbation used in the main text,
\begin{equation}
H_B=m_B\,\tau_x\otimes Q_B\otimes |N_y\rangle\langle N_y|,\qquad
Q_B=\mathrm{diag}(0,1,0).
\label{eqS:boundary_mass}
\end{equation}
projects onto this edge subspace as a Dirac mass. To leading order in $m_B$, the effective Hamiltonian becomes
\begin{equation}
H_{\mathrm{edge}}(k_x)=v k_x\sigma_x+m_{\mathrm{eff}}\sigma_y,\qquad m_{\mathrm{eff}}=\alpha m_B+O(m_B^2),
\label{eqS:effective_edge_dirac}
\end{equation}
where the coefficient $\alpha$ depends on the microscopic edge wave functions and on the sublattice content selected by $Q_B$.

\begin{figure}[h]
\centering
\includegraphics[width=\linewidth]{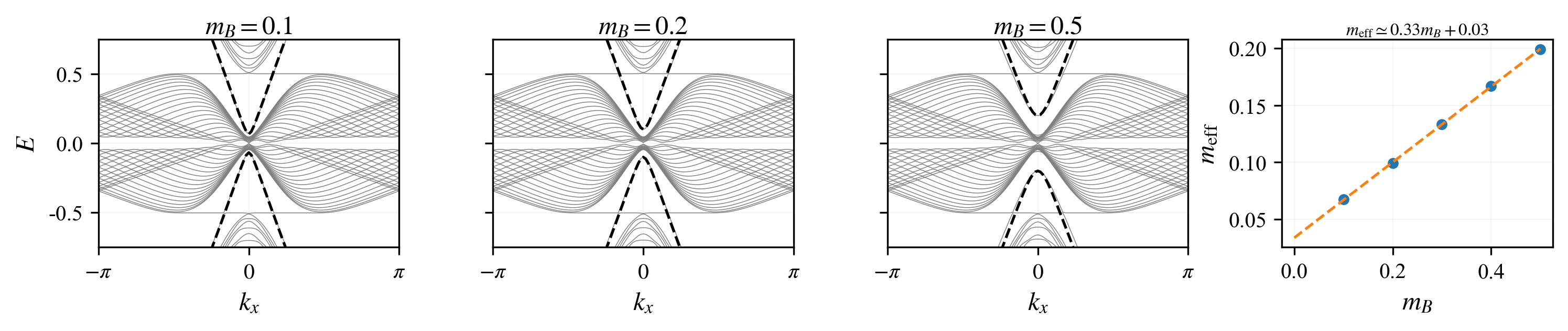}
\caption{Effective massive Dirac description of the gapped dispersive boundary branch.
(a)--(c) Low-energy ribbon spectra near $k_x=0$ for the edge-localized interlayer boundary mass in Eq.~\eqref{eqS:boundary_mass}, with $m_B=0.1,0.2,0.5$. The gray curves show the finite-strip spectrum, while the black dashed curves denote the lattice-regularized massive Dirac fit in Eq.~\eqref{eqS:dirac_fit_dispersion}.
Increasing $m_B$ opens a gap in the dispersive boundary branch, consistent with the projected massive edge Hamiltonian in Eq.~\eqref{eqS:effective_edge_dirac}.
(d) Extracted effective Dirac mass $m_{\mathrm{eff}}=\Delta_{\mathrm{edge}}(k_x=0)/2$ as a function of $m_B$. The approximately linear behavior confirms that the edge-localized interlayer perturbation generates an effective edge mass, $m_{\mathrm{eff}}\propto m_B$, in the dispersive boundary sector.}
\label{figS:boundary_dirac_mass_fit}
\end{figure}

Figure~\ref{figS:boundary_dirac_mass_fit} shows that the numerical boundary spectrum is well captured by the lattice-regularized massive Dirac dispersion
\begin{equation}
E_\pm(k_x)\simeq\pm\sqrt{\left[2\sin\left(\frac{k_x}{2}\right)\right]^2+m_{\mathrm{eff}}^2}.
\label{eqS:dirac_fit_dispersion}
\end{equation}
The extracted effective mass grows approximately linearly with the boundary-mass control parameter,
\begin{equation}
m_{\mathrm{eff}}\simeq0.33m_B+0.03,
\end{equation}
where the small offset ($\approx0.03$) is a finite-size effect.
This confirms that the edge-localized interlayer perturbation does not merely shift the edge spectrum. Instead, it acts after projection as a boundary mass and opens a genuine Dirac gap in the dispersive counter-propagating boundary sector. The edge flat band localized on the opposite boundary is not described by this projected two-state Dirac theory and remains essentially unaffected by the same edge-localized perturbation.

We now compute the continuum winding contribution of the massive edge Hamiltonian. After rescaling momentum so that $v=1$, the effective Hamiltonian is
\begin{equation}
H(k)=k\sigma_x+m\sigma_y=
\begin{pmatrix}
0&k-im\\
k+im&0
\end{pmatrix}.
\end{equation}
Writing
\begin{equation}
q(k)=k+im=E(k)e^{i\phi(k)},\qquad E(k)=\sqrt{k^2+m^2},
\end{equation}
the Hamiltonian takes the off-diagonal form
\begin{equation}
H(k)=
\begin{pmatrix}
0&q^*(k)\\
q(k)&0
\end{pmatrix}.
\end{equation}
A normalized eigenstate of the lower band $E_-(k)=-E(k)$ is
\begin{equation}
|u_-(k)\rangle=\frac{1}{\sqrt{2}}
\begin{pmatrix}
-e^{-i\phi(k)}\\
1
\end{pmatrix}.
\end{equation}
The Berry connection of this continuum lower band is
\begin{equation}
A(k)=i\langle u_-(k)|\partial_k u_-(k)\rangle=\frac{1}{2}\partial_k\phi(k).
\end{equation}
Since
\begin{equation}
\partial_k\phi(k)=\mathrm{Im}\left[\frac{\partial_k q(k)}{q(k)}\right]
=\mathrm{Im}\left[\frac{1}{k+im}\right]
=-\frac{m}{k^2+m^2},
\end{equation}
we obtain
\begin{equation}
A(k)=-\frac{1}{2}\frac{m}{k^2+m^2}.
\end{equation}
The Berry phase accumulated over the full continuum momentum line is therefore
\begin{equation}
\gamma=\int_{-\infty}^{\infty}A(k)\,dk
=-\frac{1}{2}\int_{-\infty}^{\infty}\frac{m}{k^2+m^2}\,dk
=-\frac{\pi}{2}\mathrm{sgn}(m).
\end{equation}
Equivalently, defining the continuum winding contribution by $\nu_{\mathrm{edge}}=\gamma/\pi$, one finds
\begin{equation}
\nu_{\mathrm{edge}}=-\frac{1}{2}\mathrm{sgn}(m).
\end{equation}
This half-integer value should be understood as the anomalous continuum contribution of a single massive Dirac edge fermion. It is not, by itself, a standalone integer winding number of the full lattice problem over a closed Brillouin zone. In the full lattice Hamiltonian family, the ultraviolet regularization and the remaining bands complete the total winding structure. Nevertheless, the half-integer continuum contribution is a useful diagnostic of the fact that the edge-localized interlayer coupling acts as a Dirac mass for one boundary sector, while leaving the edge flat sector on the opposite boundary essentially decoupled.

\section{Experimental considerations}

The model discussed in the main text should be viewed primarily as a target tight-binding Hamiltonian for engineered platforms rather than as an immediate material-specific proposal. A minimal experimental implementation requires four ingredients. First, one needs the Lieb-lattice connectivity with tunable nearest-neighbor hoppings. Second, one needs a two-component layer or pseudo-spin degree of freedom, with an interlayer coupling playing the role of $t_\perp$. Third, the imaginary $A$-$C$ coupling and the opposite-sign intralayer OAM-dependent coupling require controllable Peierls phases or synthetic gauge fields. Fourth, the boundary mass $H_B$ requires a spatially selective interlayer coupling acting only near one open boundary and primarily on the $B$ sublattice.

Photonic lattices provide a direct route to the spectral and boundary-state aspects of the model. Photonic Lieb lattices have already been realized in waveguide arrays, where the propagation coordinate plays the role of time and the coupled-mode equations reproduce a tight-binding Hamiltonian~\cite{Mukherjee2015PhotonicLieb}. In such a setting, the two layers of our model could be encoded either by two coupled waveguide arrays, two polarization sectors, or synthetic frequency modes. The interlayer hopping $t_\perp$ is then controlled by evanescent coupling between the two sectors, while complex hopping phases can be engineered through helical waveguides, dynamic modulation, or synthetic gauge-field schemes~\cite{Fang2012PhotonGauge,dutt2019experimental,dutt2020single}. The boundary mass in Eq.~(9) in the main text is especially natural in photonics, since one may add or remove selected couplers at a chosen edge. The main observables would be the propagation of wave packets launched near the two opposite edges, the bending of the initially flat edge band under OAM-dependent coupling, and the opening of a Dirac gap on only one boundary when the edge-localized interlayer coupling is switched on.

Topolectrical circuits offer an even more flexible platform for implementing the graph structure of the Hamiltonian. Circuit Laplacians can realize tight-binding matrices with designed positive, negative, and complex effective couplings by using capacitors, inductors, operational amplifiers, or impedance-converter elements. In this language, the two layers correspond to two copies of the Lieb circuit, $t_\perp$ corresponds to vertical circuit connections between the copies, and the boundary mass corresponds to a small number of additional links connecting only the boundary $B$ nodes. Since voltage profiles and impedance resonances are spatially resolved, this platform is well-suited for testing the asymmetric boundary response predicted in Figs.~3 and 4. Topological Lieb-lattice circuit models and flat-band/topological edge phenomena have already been simulated experimentally in related circuit networks~\cite{Zhu2018TopoLiebCircuit,Zhou2023CircuitLieb}.

Cold atoms provide a complementary route to the bulk topological diagnostic. Optical Lieb lattices have been realized with ultracold atoms~\cite{Taie2015OpticalLieb}, and schemes for Abelian and non-Abelian gauge fields on Lieb lattices have been proposed using Raman-assisted tunneling~\cite{Goldman2011LiebColdAtoms}. In this context, the layer pseudo-spin may be encoded by two hyperfine states, two optical planes, or two synthetic internal states. The opposite-sign OAM-dependent coupling in Eq.~(5) in the main text corresponds to state-dependent Peierls phases. The pseudo-spin Chern number is not a quantized transport coefficient in the metallic regime, but its projected-spin structure can be probed through spin-resolved Berry curvature or linear-response measurements, in analogy with existing measurements of spin Chern numbers in quantum-simulated topological insulators~\cite{Lv2021SpinChernColdAtoms}. The main experimental challenge for cold atoms is implementing a sharp, sublattice-selective boundary mass, which is less straightforward than in photonic or circuit platforms.

Artificial electronic lattices and solid-state realizations are also conceptually relevant but less immediate. Electronic Lieb lattices have been assembled by positioning CO molecules on Cu(111), with scanning tunneling microscopy resolving the characteristic flat-band and dispersive-band structure~\cite{Slot2017ElectronicLieb}. Such platforms are useful for motivating the Lieb-lattice geometry, but realizing the full set of ingredients in the present model---in particular, the layer-resolved opposite OAM-dependent coupling and the edge-selective interlayer mass---would require a much higher degree of microscopic control than is presently assumed in our minimal theory.

Finally, we stress the distinction between the experimental signatures of the semimetallic and metallic regimes. In the regime with $\Delta_{\mathrm{dir}}>0$ and $\Delta_{\mathrm{ind}}<0$, the system is metallic in the global spectral sense, but the occupied subspace remains locally separated in momentum space. Therefore, the relevant experimental signatures are not a quantized DC Hall or spin-Hall conductance. Instead, one should look for: (i) a negative indirect gap together with a finite direct separation between the lower and upper manifolds; (ii) persistence of a layer-resolved projected-spin gap associated with $P\tau_zP$; and (iii) boundary modes consistent with the direct-gap-protected pseudo-spin topology. In photonic and circuit platforms, where the full eigenmode profiles can be reconstructed, the projected pseudo-spin decomposition and the corresponding boundary localization can be tested directly from the measured mode amplitudes. In all platforms, the direct gap and the projected pseudo-spin gap must remain larger than the relevant disorder, loss, and finite-size broadening scales for the topological-metal interpretation to be experimentally meaningful.

\end{document}